\documentclass[10pt]{article}
\usepackage{amssymb,amsmath, psfrag, epsfig}

\def\ov{\overline}
 
\def\b{\noindent}  

\def\nh{\noindent\hangindent=1 true cm \hangafter = 1}

\def\ov{\overline}
\def\be{\begin{equation}}
\def\ee{\end{equation}} 
\def\ac{${\rm Ca_v1.3}$ }

\def\d{${\rm Ca_v1.4}$}
\def\CA{${\rm Ca^{2+}}$ }
\def\IA{${\rm I_A}$ }
\def\IAN{${\rm I_A}$}

\def\CAN{${\rm Ca^{2+}}$}

\begin{document}

\title{Properties of \IA in a serotonergic neuron
 of the dorsal raphe nucleus\\
\   \\
{\normalsize
 Nicholas J. Penington$^{1,2}$, Henry C. Tuckwell $^{3\dagger}$ \\   \
\  \\ 
$^1$ Department of Physiology and Pharmacology,\\
$^2$ Program in Neural and Behavioral Science and Robert F. Furchgott
Center for Neural and Behavioral Science \\
State University of New York,
Downstate Medical Center,\\
Box 29, 450 Clarkson Avenue, Brooklyn, NY 11203-2098, USA\\
 \              \\
$^3$ Max Planck Institute for Mathematics in the Sciences\\
Inselstr. 22, 04103 Leipzig, Germany\\
\     \\
$^{\dagger}$ {\it Corresponding author}: tuckwell@mis.mpg.de}  }

\maketitle
\noindent {\it Abbreviated title:}  Estimation of activation function for \IA in DRN

%

\begin{abstract}
Voltage clamp data were analyzed in order to 
characterize the properties of the fast potassium transient current \IA
 for a serotonergic neuron of the rat 
dorsal raphe nucleus (DRN). We obtain maximal conductance, time constants
of activation and inactivation,  and 
 the steady state activation and inactivation functions $m_{\infty}$ and
 $h_{\infty}$, as Boltzmann curves, defined by  half-activation
potentials and slope factors. We employ a novel method to
 accurately 
obtain the activation function and compare the results with those
obtained by other methods. The  form of
 \IA  is estimated as  $\ov{ g} (V-V_{rev}) m^4h$, with
$\ov{ g}=20.5$ nS. For activation,
the half-activation 
potential is $V_a=-52.5$ mV with slope factor $k_a=16.5$  mV, whereas
 for inactivation the corresponding quantities are -91.5 mV and 
-9.3 mV. 
We discuss the results in terms of  the corresponding properties of \IA in
other cell types and their possible relevance to pacemaking activity
in 5-HT cells of the DRN.

\end{abstract}


\section{Introduction}
Serotonergic neurons in the DRN,
which extensively innervate most brain regions,  
have a large influence on many aspects of
behavior, including sleep-wake cycles,
mood and impulsivity (Liu et al, 2004; Miyazaki et al., 2011). 
The firing patterns of these cells have been much 
studied and many properties of the ionic currents
underlying action potentials have been investigated
 (Aghajanian, 1985;  Segal, 1985; Burlhis and Aghajanian, 1987; Penington et al., 1991, 
1992; Penington and Fox, 1995). 
In order to quantitatively analyze 
action potential generation in DRN 5-HT neurons, 
it is desirable to have accurate  
knowledge of the activation and inactivation properties of 
the various ion currents for some of which there is relatively little or no data available.
 Here we report results for  
the fast transient
potassium current  \IA 
which  plays an
important role in determining the cell's firing rate.

Mathematical modeling of electrophysiological 
dynamics has been pursued for many different 
 nerve and muscle cell types.
Some well known neuronal examples are thalamic relay cells
(Huguenard and McCormick, 1992; Destexhe et al., 1998;  
Rhodes and Llinas, 2005), dopaminergic cells 
(Komendantov et al., 2004; Putzier et al., 2009; 
Kuznetsova et al., 2010), hippocampal
pyramidal cells (Traub et al., 1991; Migliore et al., 1995; 
Poirazi et al., 2003; Xu and Clancy, 2008) and neocortical
pyramidal cells (Destexhe et al., 2001;
Traub et al, 2003; Yu et al., 2008).  Cardiac myocytes have
also been the subject of numerous computational 
modeling studies with similar structure and equivalent complexity to that of 
neurons (Faber et al., 2007; Williams et al., 2010).

The methods employed by  Hodgkin
 and Huxley (1952) to describe 
mathematically the time (and space) course of nerve
membrane potential have, for the most part,  endured to the present day.  
An integral component of their model consists of 
differential equations for activation and (if present)
inactivation variables, generically represented in the
non-spatial models as $m(t, V)$ and $h(t, V)$, respectively, where
$t$ is time and $V$ is membrane potential. 
These equations are 
\be  \frac{dm}{dt}=  \frac{ m_{\infty}-m} { \tau_m},  \hskip .2 in \frac{dh}{dt}= \frac{ h_{\infty}-h} { \tau_h},  \ee
where $m_\infty(V)$ and $h_\infty(V)$   are steady state values  and  $\tau_m$
and $\tau_h$ are time constants 
which often depend on $V$.
For voltage-gated ion channels the current is often assumed to be
\be I=\ov{ g} (V-V_{rev}) m^ph \ee
where $\ov{ g}$ is the maximal conductance,   $V_{rev}$ is the reversal potential for the ion species
under consideration and  
 $p$ is (usually) a small non-negative integer between 1 and 4.
The inactivation is invariably to the power one, as in (2). 
Although these equations  have solutions which can only 
yield  approximations to actual currents, it is nevertheless
desirable to have accurate forms 
for the steady state activation and inactivation 
functions $m_\infty(V)$ and $h_\infty(V)$, the time constants
$\tau_m(V)$ and $\tau_h(V)$, and the remaining parameters $\ov{ g}$ and $V_{rev}$. 
Our concern in this article is to describe and illustrate a novel and 
straighforward yet accurate method of
estimating these quantities from voltage-clamp data.

\section{Results}
In order to isolate \IAN, currents for activation and inactivation protocols 
were obtained in 20 mM TEA,  as 
described in Section 4.1, for several identified serotonergic neurons of the rat
dorsal raphe nucleus. In many cells the outward currents were clearly a mixture
of components with different dynamical properties. In some cells, however,
 the currents decayed smoothly from an early peak 
to near zero and were apparently a manifestation of a single channel type,
presumed to be uncontaminated
\IAN. Thus neither CoCl$_2$ nor CdCl$_2$ was employed.
One cell in particular, DR5, with an apparently pure set of current traces was singled out
for analysis. Its activation current traces are shown in Figure 1.  
Results for other cells with composite outward currents will be analyzed 
in future articles. 

\centerline{\includegraphics[width=5.1in]{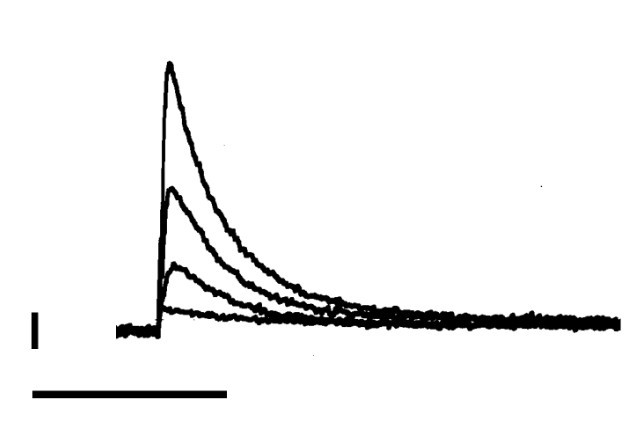}}
\begin{figure}[h]
\caption{Activation current traces under voltage clamp for cell DR5 in 20 mM TEA.
Time marker, 100 ms, Current marker, 100 pA. Voltage stepped from -120
mV to -60, -50, -40, -30 and -20 mV. For experimental protocol see Section 4.1.}
\label{}
\end{figure}

The current traces of Figure 1  were obtained in digitized form and  
the time constants $\tau_m$ and   $\tau_h$ and the value of $p$
were estimated 
by the best fitting of the current traces to the analytical formula (6).
Only two of these three parameters can be independently chosen by virtue
of the constraint given by  formula (7) for  the time of occurrence $t_{max}$
of the maximum current, which rearranges
to
\be \tau_h  =  \frac{ \tau_m  }{p  } \bigg( e^{   \frac{  t_{max} }{  \tau_m }    -1      }   \bigg)  \ee
The current was required to pass through the maximum point
at $(t_{max}, I_{max})$ and one other point on the trace, called $(t_2, I_2)$. 
The value of $p=4$ was found to give excellent fits to the current traces. 
 An example of the resulting curve with best fitting time constants 
$\tau_m$ and    $\tau_h$
 is shown in Figure 2 along with the experimental current for a step from -120 mV to -20 mV.

Table 1
summarizes the data and estimates of time constants for cell DR5, all of
which quantities are required to estimate the three parameters 
$\ov{ g}$ (conductance), $V_a$ and $k_a$ (Boltzmann
parameters for activation). In this table \IAN$_{,max}$ is the maximal experimental
current over time.

\centerline{\includegraphics[width=4.5in]{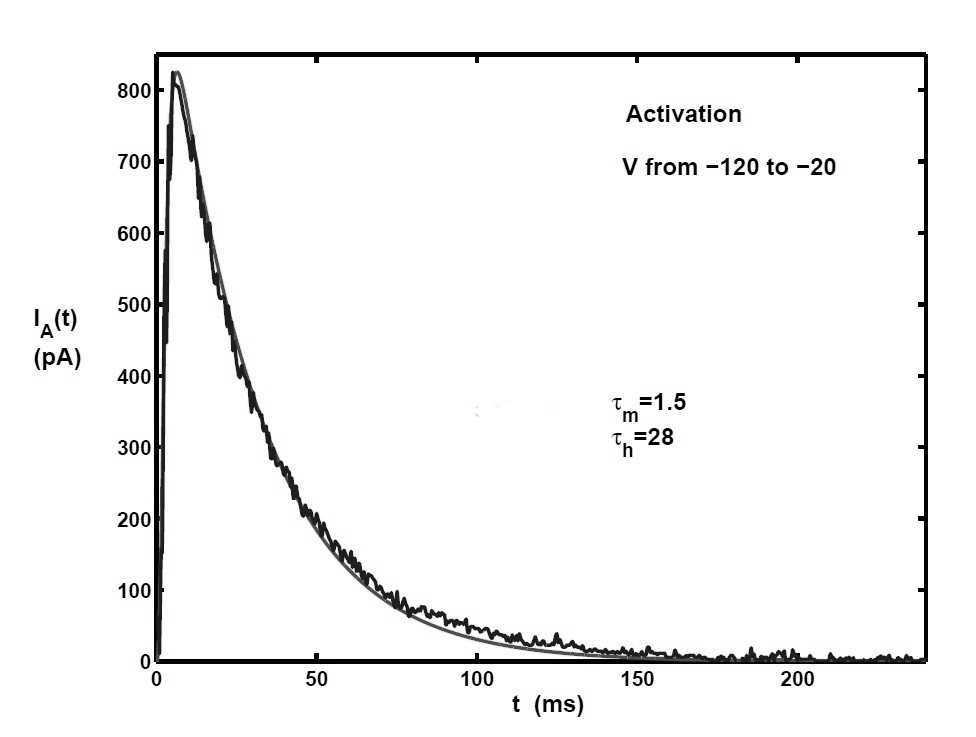}}
\begin{figure}[!h]
\caption{An example of the best fitting curve for the calculated time-dependent current \IAN$(t)$ (smooth curve), as given in Equation
(6) with $p=4$,  to an experimental trace for cell DR5 with calculated time constants
for activation  $\tau_m=1.5$ ms, and inactivation $\tau_h=28$ ms. 
Voltage stepped from -120 mV to -20 mV.}
\label{}
\end{figure}

\begin{center}
\begin{table}[h]

    \caption{Maximal currents and estimated time constants in}
\centerline{activation experiments for \IA in cell DR5}
\smallskip
\begin{center}
\begin{tabular}{cccc}
\hline
Step from -120 mV &  \IAN$_{,max}$ (pA) & $\tau_m$ (ms) & $\tau_h$ (ms)  \\
to &   &  &   \\
\hline
-20 mV  & 825.4 & 1.5 & 28.0  \\
-30 mV  & 431.7 & 1.5 & 28.0  \\
-40 mV  & 171.5 & 2.4 & 21.7  \\
-50 mV  & 22.2 & - & -  \\
-60 mV  & 0 &  - &  -  \\
\hline
\end{tabular}
\end{center}
\end{table}
\end{center}

The results obtained by nonlinear least-squares fitting (Section 4.2.2) with correction factor 
as in Equation (8) and following but without making any assumption
about $V^*$,  (called method A), are that for cell DR5
the whole cell maximal conductance for the \IA channels is $\ov{g} = 20.5$ nS,
the half-activation potential is $V_a =-52.5$ mV and the corresponding slope factor is
$k_a = 16.5$ mV (see also Table 2).

As a check on the estimated parameter values by method A, the maximal
currents with steps from $V_0= -120$ mV to  $V_1$ =-60, -50, -40, -30 and -20 mV
  were calculated by formula (8) using the values 
$\ov{g} = 20.5$ nS, $V_a=-52.5$ mV, $k_a=16.5$ mV, $p=4$, $V_{rev}=-105$ mV
and the time constants given in Table 1. The maximal currents so determined are
plotted against $V_1$  in Figure 3 where it can be seen that there is
excellent agreement with the experimental
values. Also shown in the inset of the Figure is the activation function 
raised to the power 4, a so called 4th order Boltzmann.

 \centerline{\includegraphics[width=5in]{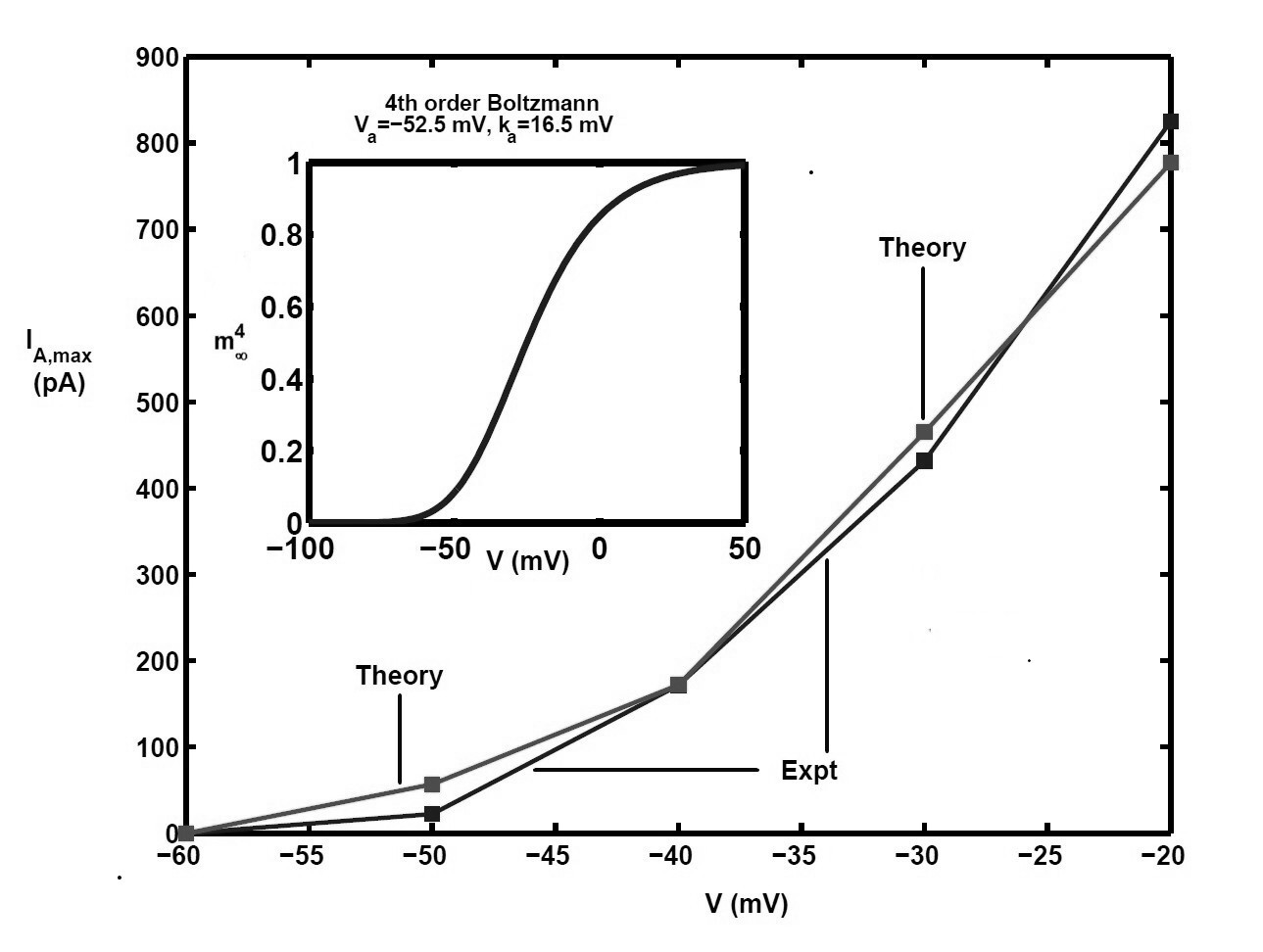}}
\begin{figure}[ht]
\caption{Predicted and experimental values of $I_{max}$ in activation
protocols stepped up from -120 mV to various clamp
voltages, in mV, -60, -50, -40, -30 and -20, for cell DR5 with TEA 20 mM. 
The inset shows the best fitting 4th order Boltzmann.}
\label{}
\end{figure}


It is interesting to compare the values of the activation function
so obtained with those obtained by three other procedures. 
For method B, in which it is assumed that 
 $V^*=-20$ mV so that $m_{\infty}(-20)=1$ (see note after Equation (11)), which implies that
Equation (12) gives an estimate of $\ov{g}$, the values $\ov{g}=12.91$ nS, 
$V_a=-54.7$ mV and $k_a=12.52$ mV are obtained.

  In method C, the value of
$V^*$ is not fixed but a best-fitting Boltzmann is obtained by least sqares, 
  which gives 
$V_a=-47.0$ mV and $k_a=10.2$ mV.  It is also of interest to compare these sets of results with
those by the usual method, called method D,  using the raw $G/G_{max}$ values
which assumes that $V^*=-20$ mV.   This gives $V_a=-48.2$ mV and $k_a=11.78$ mV. 

  It can be seen from the data in Table 2 that there are
considerable discrepancies between the results obtained by the
various methods.  Method A gives the most accurate estimates. 
For the other three methods, relative errors in the conductance are between -37.1 \% and -52.6 \% 
 relative errors in $V_a$ are from  -10.5 \% to +4.2 \%
and those in $k_a$ are from  -24.1 \%  to  -38.2 \%.  Such errors
can clearly lead to significant errors in the computed ionic currents.

\begin{center}
\begin{table}[ht]

    \caption{Estimated activation function and conductance of \IAN}
\centerline{for cell DR5, obtained by four methods as described in the }
\centerline{text: \% errors relative to method A given in brackets}
\smallskip
\begin{center}
\begin{tabular}{ccccc}
\hline
Method  $\rightarrow$ &  A &  B & C & D   \\
\hline
$\ov{g}$ (nS) & 20.5 & 12.9 (-37.1)   &  12.4 (-39.5)  &  9.71 (-52.6)  \\
\hline
 $V_a$ (mV) & -52.5 &   -54.7 (+4.2)  & -47 (-10.5)  & -48.2 (-8.2)  \\
\hline
 $k_a$ (mV) & 16.5 & 12.52 (-24.1) &   10.2  (-38.2) & 11.78 (-28.6)  \\
\hline
\end{tabular}
\end{center}
\end{table}
\end{center}

\subsection{Inactivation experiments}

For the inactivation voltage-clamp experiments, whose results are not shown,
 the
maximum currents $I_{max}(V_0,V_1)$ at various starting potentials 
 $V_0$ are sufficient to estimate the relative values of
the inactivation function because
all factors on the right hand side of (17) except $h_{\infty}(V_0)$ 
do not depend on $V_0$. Assuming that $h_{\infty}(-120)=1$,
the best fitting steady state inactivation function was 
 \be h_{\infty}(V) = \frac{1 }{   1 +  e^{(V+91.5)/9.3}}.  \ee
\subsection{Graphical summary}
The activation functions obtained by methods A-D  and the inactivation function 
for the transient potassium
current \IA in cell DR5 are drawn in the top panel of Figure 4. Also shown, in the
lower panel, 
is the activation function to power 4, which reflects more 
closely the contribution of activation to
current amplitudes at various membrane potentials.

 \centerline{\includegraphics[width=5in]{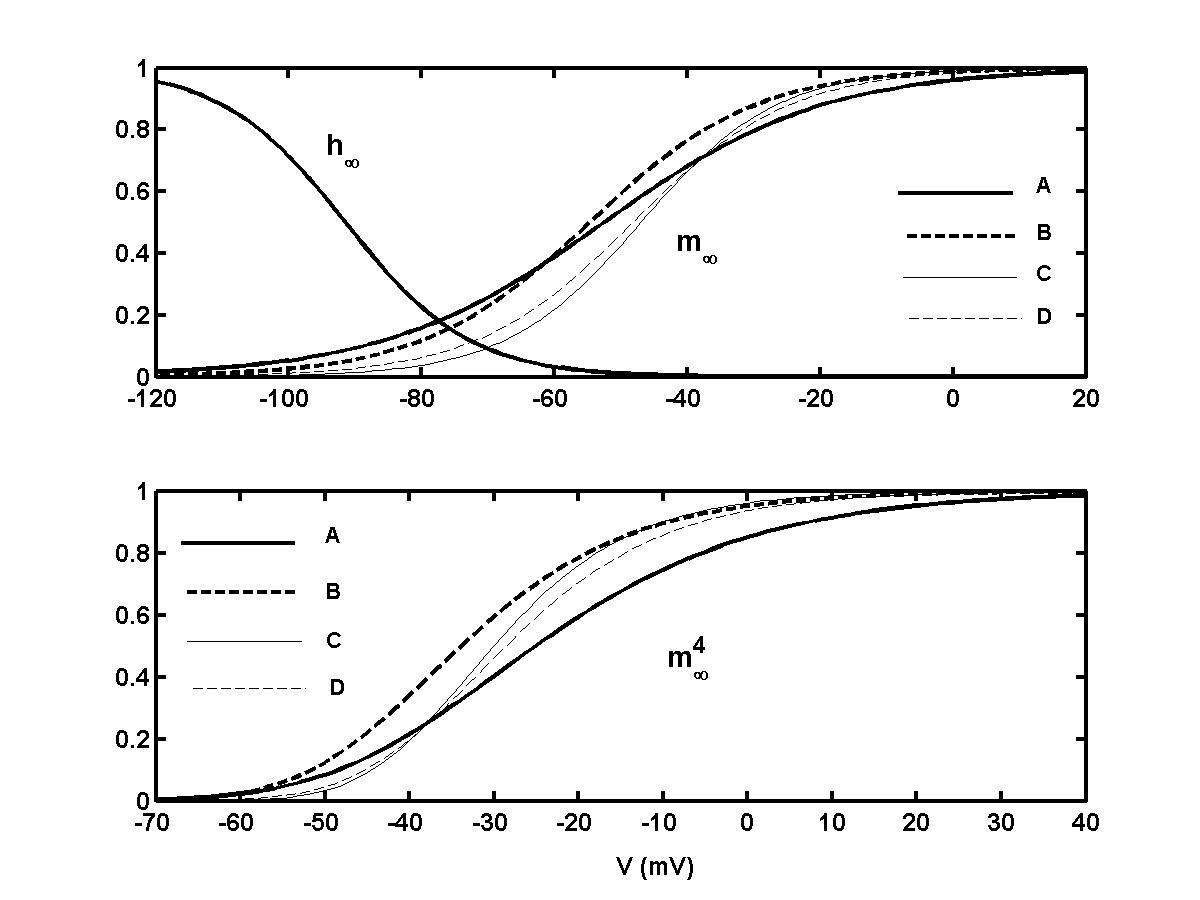}}
\begin{figure}[ht]
\caption{{\it Top part.} Steady state inactivation function  and steady state activation function 
obtained by the methods A-D 
as explained in the text for \IA in cell
DR5 obtained from voltage-clamp data by the methods described.
The result for method A is the most accurate estimate of $m_{\infty}.$
{\it Lower part.}
The  activation function  to the power 4 is shown for the methods A-D.}
\label{}
\end{figure}

\section{Discussion}

There have been many investigations of the 
properties of the ionic currents involved in the firing of serotonergic 
neurons of the dorsal raphe nucleus (Aghajanian, 1985;
Segal, 1985; Burlhis and Aghajanian, 1987; Penington et al., 1991, 
1992;
Penington and Fox, 1995; Chen et al., 2002). 
However, the basis of apparent pacemaker-like  activity 
in these cells is not fully understood.  It was
claimed that noradrenergic input is required (Burlhis and
Aghajanian, 1987) for firing in some cells  and there have been
conflicting ideas about the role of \IA (Burlhis and
Aghajanian, 1987; Segal, 1985).  No reports on the properties
of \IA in these cells have previously appeared.

 In the first quantitative study of \IAN, Connor and Stevens (1971)
reported an $m^4h$ form for the current in  a single gastropod neuron 
and found quite long time  constants
for activation (12 ms) and inactivation (235 ms). The   $V_{1/2}$ value for inactivation
was -68 mV, with a resting potential of -40 mV.  It was also found that
\IA was important mainly in the middle and end of the inter-spike interval (ISI, see below).
Concerning the method of analyzing voltage-clamp data for activation,
in their pioneering work on squid axon,  for large depolarizations
 Hodgkin and Huxley (1952) 
used the general Equation (5) for the time-dependent sodium
current converted to a conductance 
$$g(t)=\ov{ g} [m_{\infty}(V_1)
(1- e^{-t/\tau_m})]^3h_{\infty}(V_0) e^{-t/\tau_h} $$
to estimate $\ov{ g} [m_{\infty}(V_1)]^3h_{\infty}(V_0)$. 
Belluzzi and Sacchi (1988) also employed 
Equation (5) with simplifying assumptions for $m$ and $h$,
and $p$ as a parameter. 
Activation time constants were estimated via Equation (7) with $p=3$,
the small values of both $\tau_m$
and $\tau_h$ they obtained reflecting the relatively high temperature of the preparation.
For activation in histamine neurons, 
Greene et al. (1990) found  a fast and slow component
of \IA and reported Boltzmann parameters for 
normalized current without converting to conductances. 
Huguenard and McCormick (1992) used normalized currents 
to find inactivation parameters and time-courses of conductances 
(as in (5)) to determine Boltzmann parameters for activation.
Bekkers (2000) gave two sets of results for activation,
one obtained from the empirical Boltzmann fit to the peak \IA conductance curve,
 and the other obtained from the  time-dependent \IA conductance curves.
We have
pursued a new approach, as outlined in the last section, which 
utilizes the maxima of the currents at various clamp voltages and
employed a correction factor as in Equation (8).

The results  obtained for \IA in a serotonergic cell of rat DRN 
are compared with those in some  other CNS preparations in Table 3.
It can be seen that the half-activation potentials for activation are generally
similar and that the largest variation occurs in the time constant of inactivation. 
\begin{center}
\begin{table}[ht]
    \caption{Activation and inactivation parameters for \IA in some CNS cells}
\smallskip
\begin{center}
\begin{tabular}{lccccc}
\hline
 &  DRN  &   Sympathetic &  Thalamic&  Cortical  \\
  & serotonergic$^a$ &  cervical ganglion$^b$ & relay$^c$  & pyramid$^d$ \\
\hline
Activation&  &    &     &  \\
\hline
 $V_{1/2}$ mV & -52.5 &   -58.55  & -60  &  -56.2 \\
 $k$ mV& 16.5 & 14.39  & 8.5  &  19.1  \\
 $\tau_m$ ms & 1.5-2.4 &$ < 1$ & 0.5-2.5  &0.3-8   \\
\hline
Inactivation&  &    &   &    \\
\hline
 $V_{1/2}$ mV & -91.5 &   -78  &   -78 & -81.6 \\
 $k$ mV & 9.3 & 7.3 &   6 & 6.7 \\
 $\tau_h$ ms & 21.7-28.0  & $<10$ &  12-65 & 6-8  \\
\hline
\end{tabular}
\noindent  $^a$This study; $^b$Belluzzi \& Sacchi (1988);  $^c$Huguenard \& McCormick (1992);
 $^d$Bekkers (2000).
\end{center}
\end{table}
\end{center}

In order to construct a computational
model of 
action potential generation in DRN 5-HT neurons,  such that  
the role of the various ionic components can be
properly understood, 
it is desirable to have accurate  
knowledge of the activation and inactivation properties of 
the various ion currents, including that addressed here, \IAN. 
The need for considerable accuracy is 
made the more important because several components,
including the low threshold T-type calcium and \IA operate in
overlapping ranges of potential.
In this paper we have illustrated that the often-used procedure of finding
activation functions $m_{\infty}(V_1)$ by taking ratios of current to maximal current or conductance
to maximal conductance may produce inaccurate results due to the fact
that the quantity $F_p(\gamma)$ in (8) depends on the clamp voltage and does
not cancel. Not taking this into account can lead to substantial
errors in the estimates of the  parameters $k_a$ and $V_a$.  Inactivation functions, however,
do not have such a complication and can be estimated in the usual way.

 L-type calcium
currents, particularly through \ac channels, have been found to sometimes play a role
in pacemaker activity in neurons and cardiac cells
 (Koschak et al., 2003; Putzier et al., 2009;
 Marcantoni et al., 2010).    
However, in DRN 5-HT cells, there is only a small (about 4\% of total $I_{Ca}$)
 contribution from L-type calcium currents (Penington et al., 1991)
so it is yet to be determined what role these play in the pacemaking activity. 
These latter results were, however, obtained in dissociated cells. 
Burlhis and Aghajanian (1987) hypothesized that the low threshold
calcium T-type current played a key role in pacemaking and Segal (1985) proposed
that \IA was also important.  However, considering the fact that the resting potential
for these cells is about -60 mV and that threshold for spiking is about -50 mV,
and judging by the curves for  $m^4_{\infty}(V)$    and   $h_{\infty}(V)$   in Figure 4, it is likely that 
\IA is only important during the spike itself and not between the commencement
of the afterhyperpolarization and the next spike, probably implying a lesser
role for \IA in pacemaking as it tends to be switched off during the greater part
of the ISI. We will explore 
by means of a computational model  
the ionic basis of the mechanisms of pacemaking in  DRN 5-HT cells 
in future articles. 

Finally we note that the \IA results obtained here for $m_{\infty}$ and the time constants
 bear a resemblance
to those given in Gutman et al. (2005) for the transient A-type channels
$K_v4.1$ ( $V_a=-47.9$ mV, $k_a = 24$ mV,  $\tau_h=22$ ms, and two other 
 $\tau_h$ values).  This contrasts with the finding by Bekkers (2000) who suggested
that in rat layer 5 cortical pyramidal cells, 
 \IA  was carried by $K_v4.2$ channels, which accounts for their smaller time constants
of inactivation (Gutman et al., 2005). 

\section{Materials and Methods}
\subsection{Experimental}
Data of the kind analyzed in this paper were obtained from 2 month-old male
 Sprague-Dawley rats that were anesthetized with halothane and then
 decapitated with a small animal guillotine in accordance with our local
 Animal Care and Use Committee regulations. Three coronal slices 
(500 $\mu$m) through the brain stem at the level of the DRN were
 prepared using a “vibroslice” in a
 manner that has previously been described (Yao et al., 2010).
  Pieces of tissue containing the DRN were then incubated in a 
PIPES buffer solution containing 0.2 mg/mL trypsin (Sigma Type XI)
 under pure oxygen for 120 min. The tissue was then triturated in
 Dulbecco's modified Eagle's medium to yield individual
 neuronal cell bodies with vestigial processes. Recording was carried
 out at room temperature, 20$^\circ$C-25$^\circ$C. Cells recorded
 from had diameters of about 20 $\mu$m. Steinbusch et al. (1981)
 showed using immunohistochemistry that most of the cells
 in a thin raphe slice with a soma diameter greater than 20 $\mu$m 
contain 5-HT while the smaller cells were largely GABAergic 
interneurons.  Using a method
 similar to that of Yamamoto et al. (1981), we also found that the proportion
 of isolated cells
 with diameters larger than 20 $\mu$m that stain for 5-HT, was greater than 85\%, and 
a similar percentage responded to serotonin.

The cells were voltage-clamped using a switching voltage
 clamp amplifier (Axoclamp 2A) and a single patch pipette in
 the whole-cell configuration (Hamill et al., 1981). Pipettes 
pulled from thick-walled borosilicate glass (resistance from 5-7 M$\Omega$)
 allowed a switching frequency of 8-13 KHz with a 30\% duty cycle. 
The seal resistance measured by the voltage response to a 50 pA step of
 current was often greater than 5 G$\Omega$). The voltage-clamp data were filtered at
 1 KHz and digitized for storage at 16 bits, 
 or experiments were run online with a PC and a CED 1401 interface.

The external saline was designed to isolate potassium currents and 
contained: outside the cell TTX  0.2  $\mu$M and in mM NaCl 147, KCl 2.5,
 CaCl$_2$ 2, MgCl$_2$ 2, Glucose 10, HEPES 20, ph 7.3.
 Inside: K$^+$ Gluconate 84, MgATP 2, KCl 38, EGTA-KOH 11, 
HEPES 10, CaCl$_2$ 1, pH 7.3. The total [K$^+$] inside was 
155mM (since 33 mM is added to make the salt of EGTA).
 The osmolarity was adjusted with sucrose so that the pipette 
solution was 10 mOsm hypoosmotic to the bath solution. 
 Drugs were either dissolved in the extracellular 
solution and added to the perfusate or applied by diffusion
 from a patch pipette (tip 15 $\mu$m) lowered close to the cell (50 $\mu$m
 and then removed from the bath). As a control, the same procedure
was carried out without addition of the drug to the application pipette;
 this did not affect \IAN.  In some cells,  to investigate any influence of \CA  
currents on \IA  either 2 mM CoCl$_2$ or 20 $\mu$M CdCl$_2$
 was added to the bath by a pipette placed near the cell. 
 This sometimes changed the magnitude of $I_A$ but had little effect
on its time course.

\subsection{Theory}

 Suppose that in  a voltage clamp experiment
the voltage is, after equilibration at a voltage $V_0$, suddenly clamped at
the new voltage $V_1$.  Then, assuming a current of the form of Equation (2),
 according to the standard Hodgkin-Huxley  (1952) theory, 
 the current at time $t$ after the switch to $V_1$ is
\be  I(t; V_0, V_1) =\ov{ g} (V_1-V_{rev})
[m_1 - (m_1-m_0) e^{-t/\tau_m}] ^p  [h_1 - (h_1-h_0) e^{-t/\tau_h}], \ee
where we have employed the abbreviations
$m_0= m_{\infty}(V_0),  m_1= m_{\infty}(V_1)$ 
for the activation,  $h_0= h_{\infty}(V_0),  h_1= h_{\infty}(V_1)$
for the inactivation and 
$ \tau_m = \tau_m(V_1),  \tau_h = \tau_h(V_1)$ for the
time constants.

\subsubsection{Activation experiments}
In activation eperiments there is a step up from a relatively hyperpolarized
state $V_0$ to several more depolarized states $V_1$ so that one usually
assumes that $h_0 = 1, h_1 = 0$ and $m_0 = 0$. 
This gives the simplified expression for the  current
\be  I(t; V_1) =\ov{ g} (V_1-V_{rev})[m_{\infty}(V_1)
(1- e^{-t/\tau_m})]^p e^{-t/\tau_h}. \ee
Finding the maximum by differentiation yields the time at which the maximum
occurs as
\be  t_{max}(V_1) =\tau_m \ln \bigg(1 +   \frac{ p\tau_h   }{  \tau_m } \bigg)  \ee
and its value as
\be  I_{max}(V_1) = \ov{ g} (V_1-V_{rev})m^p_{\infty}(V_1)F_p(\gamma)   \ee
where 
\be  F_p(\gamma) = \frac{ (p\gamma)^p }{ (1 + p\gamma)^{p + 1/ \gamma   }}    \ee 
and where  $\gamma$, which depends on $V_1$, is defined as
\be  \gamma(V_1)=\frac{\tau_h(V_1) }{\tau_m(V_1)  }.   \ee
 Since $V_1$ is here considered a variable, the time
course of the current changes as $V_1$ varies. By rearrangement the steady
state activation is given by

\be m_{\infty}(V_1) = \bigg( \frac{ I_{max}(V_1)  }{  \ov{ g} (V_1-V_{rev})F_p(\gamma) 
   } \bigg)^{\frac{1}{p}}    \ee

If it is known that for some (usually well-depolarized) state $V = V^*$ 
one
has  $m_{\infty}(V^*) = 1$ (but note that this can only ever be
approximately true as $m_{\infty}$ can only approach unity asymptotically)   then the maximal conductance
 $\ov{ g}$  can be estimated from
voltage clamp experiments from
\be \ov{ g} =  \frac{ I_{max}(V^*)  }{   (V^*-V_{rev})F_p(\gamma(V^* ))    }.     \ee
Once $\ov{ g}$ is known then (11) can be used to find $ m_{\infty}(V_1)$
for various $V_1$ because all remaining quantities, $I_{max}$,
$(V_1-V_{rev})$ and  $F_p(\gamma(V_1))$ are known.

In summary, the steps to find the activation function $m_{\infty}(V)$ are:

\b (a) From the time course of the current find the maximum current
     and its time of occurrence for a given $V_1$

\b (b) Estimate the time constants $\tau_m(V_1)$ and $\tau_h(V_1)$  and
the value of $p$.

\b (c) Hence find $\gamma(V_1)$ and  also $F_p(\gamma(V_1))$.

\b (d) Using a suitably highly depolarized state  $V = V^*$ estimate  $\ov{ g}$ according
to (12).

\b (e) Hence estimate $m_{\infty}(V)$ at various $V$ below  $V^*$. 

\b (f) Find the best fitting Boltzmann curve passing through the 
values obtained. 

However, the formulas (6)-(8) were derived on the assumption that
$h(V_1)=1$ and this will usually be valid for $V_c < V_1< V^*$,
where $V_c$ is below the half-activation potential $V_a$. This 
means that
the half-activation function  $m_{\infty}(V)$  can be obtained accurately 
for $V \ge V_a$, and thus by symmetry for $V < V_a$.  Another  
approach is as follows.
\subsubsection{$V^*$  is not known: least squares estimates}
If $V^*$ is not known or is uncertain, then 
least squares estimates can be made as follows. 
 It is assumed that 
 \be m_{\infty}(V) = \frac{1 }{   1 +  e^{-(V-V_a)/k_a}}  \ee
is the steady state activation function. Then one may estimate $\ov{ g}$, $V_a$ and $k_a$
by minimizing the sum of squares of the differences between 
experimentally obtained maxima of the current at various $V_i$ 
and the values given by  formula (8)

\subsubsection{Summary of methods A-D used to estimate the activation function}
Here is given a brief description of the 4 methods employed to
estimate the parameters of  $m_{\infty}(V)$ and  $\ov{ g}$ given in Table 2.
\smallskip \\
\noindent  {\bf Method A}\\
Having estimated the values of the time constants
$\tau_m(V_i)$ and $\tau_h(V_i)$ and the power $p$ as described in the text, 
experimental maximum currents at various $V_i$ are compared with
those obtained from formula (8)
\be  I_{max}(V_i) = \ov{ g} (V_i-V_{rev})m^p_{\infty}(V_i)F_p(\gamma)   \ee
with $\gamma$ evaluated at $V_i$. Thus the parameters
 $\ov{ g}$, $V_a$ and $k_a$ 
are estimated by least squares.  
\smallskip \\
\noindent  {\bf Method B}\\
It is assumed that $m_{\infty}(V^*)=1$, with  $V^*= -20$ mV in the present example,
so that   $\ov{ g}$ is given directly by Equation (12).  Then $m_{\infty}(V_i)$
can be obtained for various $V_i$ uising Equation (11).  A best fitting 
Boltzmann is then obtained for the various vaules of $m_{\infty}(V_i)$.
 \smallskip \\
\noindent  {\bf Method C}\\
The experimental values of the maximum currents (over time) $I_{i,max}(V_i)$ at each test voltage $V_i$ 
are divided by $V_i-V_{rev}$ to give corresponding conductances $G_i(V_i)$. 
The conductances $G_i(V_i)$ are assumed (from Equation (8) without the correction factor)
to be given by 
\be G_i(V_i) = \ov{ g}  m^p_{\infty}(V_i). \ee 
With  \be m_{\infty}(V) = \frac{1 }{   1 +  e^{-(V-V_a)/k_a}}  \ee
the parameters $\ov{ g}$, $V_a$ and $k_a$ (and possibly $p$ if not already known)
are estimated by least squares 
fitting  of the experimental $G_i$ values to the predicted ones, without any assumption
on $V^*$ (normalization). 
\smallskip \\
\noindent  {\bf Method D}\\
The same procedure is carried out as in method C, but the $G_i$ values are divided
by the maximum value $G_{max}$, assumed to occur at $V^*$ (-20 mV in the above), to give the normalized values
\be  \tilde{G_i} =  \frac{G_i }{ G_{max}}.   \ee
Then since $\tilde{G_i}(V^*)=1$
there are only two parameters  $V_a$ and $k_a$  to estimate with a best fitting Boltzmann curve.
The maximal conductance $\ov{ g} $ can be obtained as 
\be \ov{ g} = \frac{I_{max}(V^*) }{V^*-V_{rev}}.  \ee

 \subsubsection{No inactivation}
If there is no inactivation in the course of the clamp experiment then
the maxiumum current will be the asymptotic value
\be  I_{max}(V_1) = \ov{ g} (V_1-V_{rev})m^p_{\infty}(V_1).  \ee
So, with   $V^*$ as above, 
  $\ov{ g}$ is again given by (18), 
and it is straightforward to obtain the required activation function
from the remaining  $I_{max}$ measurements.
\subsubsection{Inactivation experiments}
Here steps are made from a number of relatively hyperpolarized states $V_0$
to a fixed relatively depolarized state $V_1$. It is usually assumed that $h_1 = 0$
and $m_0 = 0$,  so that the current is approximately
\be  I(t; V_0, V_1) =\ov{ g} (V_1-V_{rev})m^p_{\infty}(V_1)h_{\infty}(V_0)
(1-e^{-t/\tau_m})^p  e^{-t/\tau_h}. \ee
Here the time course of the current (but not its magnitude) is the same for
all $V_0$.

 The time of occurrence of the maximum $  t_{max}$ of $I(t; V_0; V_1)$
 is again given
by (7) but the value of the maximum is now
\be  I_{max}(V_0,V_1) = \ov{ g} (V_1-V_{rev})m^p_{\infty}(V_1)h_{\infty}(V_0)F_p(\gamma).  \ee

Assuming that  $\ov{ g}$  has already been estimated and that
$ V^*$ is such that $m_{\infty}(V^*)=1$, then the inactivation
function can be found at $V_0$ from
\be h_{\infty}(V_0) ={ I_{max}(V_0,V^*)} { \ov{ g} (V^*-V_{rev}) F_p(\gamma(V^*))} \ee
It can be seen that normalizing the currents by dividing by the maximum of the  $I_{max}$
gives a reasonable estimate of the steady state inactivation function.

\section*{Acknowledgements}  
Support from the Max Planck Institute is appreciated (HCT). 
We appreciate useful correspondence with Professor John Huguenard (Stanford University)
and Professor John Bekkers (Australian National University).

\section*{References}

\nh Aghajanian, G.K., 1985. Modulation of a transient outward current in serotonergic
 neurones by $\alpha_1$-adrenoceptors.  Nature 315, 501-503.

\nh Bekkers, J.M.,  2000. Properties of voltage-gated
potassium currents in nucleated
patches from large layer 5 cortical pyramidal neurons
of the rat. J Physiol 525.3, 593-609.

\nh Belluzzi, O., Sacchi, O., 1988. 
The interactions between potassium and sodium currents
in generating action potentials in the rat sympathetic
neurone.  J Physiol 397, 127-147.

\nh Burlhis, T.M., Aghajanian, G.K., 1987. Pacemaker potentials
 of serotonergic dorsal raphe neurons:
 contribution of a low-threshold \CA conductance. Synapse 1, 582-588.

\nh Chen, Y.,  Yao, Y., Penington, N.J., 2002. 
Effect of pertussis toxin and {\it N}-ethylmaleimide on
voltage-dependent and -independent 
calcium current modulation in serotonergic neurons. 
Neuroscience 111, 207-214.  

\nh Connor, J.A., Stevens, C.F., 1971.   
Prediction of
repetitive firing behaviour from voltage clamp
data on an isolated neurone soma.
J Physiol 213, 31-53. 

\nh Destexhe, A., Neubig, M., Ulrich, D.,  Huguenard, J., 1998. 
Dendritic low-threshold calcium currents in thalamic relay cells.
 J Neurosci 18, 3574-3588. 

\nh Destexhe, A., Rudolph, M., Fellous, J-M., Sejnowski, T.J., 2001.
Fluctuating synaptic conductances recreate {\it in vivo}-like
activity in neocortical neurons. Neuroscience 107, 13-24.

\nh Faber, G.M., Silva, J., Livshitz, L., Rudy, Y., 2007. Kinetic properties of the
 cardiac L-Type Ca$^{2+}$  Channel and its role in
myocyte electrophysiology: a theoretical investigation.  
Biophys J 92, 1522-1543.

\nh Gutman, G.A., Chandy, K.G., Grissmer, S., Lazdunski, M., Mckinnon, D., Pardo,
L.A., Robertson, G.A., Rudy, B., Sanguinetti, M.C., St\"uhmer, W., Wang, X.,
2005. International Union of Pharmacology. LIII. Nomenclature and
molecular relationships of voltage-gated potassium channels. Pharmacol
Rev 57, 473508.

\nh  Hamill, O.P., Marty, A., Neher, E., Sakmann, B., Sigworth, F..J., 1981.
 Improved patch-clamp techniques for high-resolution current 
recording from cells and cell-free membrane patches. 
Pflugers Arch 391, 85-100.

\nh Hodgkin, A.L., Huxley, A.F., 1952. A quantitative description of membrane
current and its application to conduction
and excitation in nerve.  J Physiol 117, 500-544.

\nh  Komendantov, A.O., Komendantova, O.G., Johnson, S.W., Canavier, C.C., 2004.
A modeling study suggests complementary roles for GABA$_A$ and NMDA
receptors and the SK channel in regulating the firing pattern in midbrain
dopamine neurons. J Neurophysiol 91, 346-357.

\nh Koschak, A., Reimer, D., Walter, D. et al., 2003.  \d$\alpha_1$ 
 subunits can form slowly inactivating
dihydropyridine-sensitive L-type \CA  channels lacking
\CA-dependent inactivation. J Neurosci 23, 6041-6049.

\nh Kuznetsova,  A.Y., Huertas, M.A., Kuznetsov, A.S., Paladini, C.A., Canavier, C.C., 2010.
 Regulation of firing frequency in a computational model of a
midbrain dopaminergic neuron. 
J Comp Neurosci 28, 389-403.

\nh Liu, R-J., van den Pol, A.N., Aghajanian, G.K., 2004. 
Hypocretins (orexins) regulate serotonin neurons in the dorsal
raphe nucleus by excitatory direct and inhibitory indirect actions. 
J Neurosci 22,  9453-9464.

\nh Marcantoni, A., Vandael, D.H.F., Mahapatra, S., Carabelli, V., Sinnegger-Brauns, M.J., 
Striessnig, J., Carbone, E., 2010.   Loss of \ac channels reveals the critical role of L-type
and BK channel coupling in pacemaking mouse adrenal
chromaffin cells. J Neurosci 30, 491-504.

\nh McCormick, D.A., Huguenard, J.R., 1992.  A model of the
 electrophysiological properties of 
thalamocortical relay neurons. J Neurophysiol 68, 1384-1400.

\nh Migliore, M., Cook, E.P., Jaffe, D.B., Turner, D.A., Johnston, D., 1995.
Computer simulations of morphologically reconstructed CA3 hippocampal neurons.
J Neurophysiol 73, 1157-1168.

\nh Miyazaki, K., Miyazaki, K.W., Doya, K., 2011. Activation of dorsal raphe
 serotonin neurons underlies
waiting for delayed rewards. J Neurosci 31, 469-479.

\nh Penington, N.J., Fox, A.P., 1995.  Toxin-insensitive Ca current
 in dorsal raphe neurons. J Neurosci 15, 5719-5726.

\nh Penington, N.J.,  Kelly, J.S.,  Fox, A.P., 1991. A study of the mechanism of Ca$^{2+}$ current 
inhibition produced by
serotonin in rat dorsal raphe neurons.  J Neurosci I7, 3594-3609.

\nh Penington, N.J.,  Kelly, J.S.,  Fox, A.P., 1992. 
Action potential waveforms reveal simultaneous changes
 in I$_{\text{Ca}}$ and I$_{\text{K}}$
 produced by 5-HT in rat dorsal raphe neurons.  
Proc R Soc Lond. B 248, 171-179.

 \nh Penington, N.J., Fox, A.P.,  (1995) Toxin-insensitive Ca Current in dorsal raphe
neurons. J Neurosci 15, 5719-5725.

\nh Poirazi, P., Brannon, T., Mel, B.W., 2003.   Arithmetic of subthreshold synaptic
summation in a model CA1 pyramidal cell. Neuron 37, 977-987. 

\nh Putzier, I., Kullmann, P.H.M., Horn, J.P., Levitan, E.S., 2009. 
Ca$_v$1.3 channel voltage dependence, not Ca$^{2+}$ selectivity,
drives pacemaker activity and amplifies bursts in nigral
dopamine neurons. J Neurosci 29, 15414-15419.

\nh Rhodes, P.A., Llin\'as, R., 2005.  A model of thalamocortical relay cells. 
J Physiol 565, 765-781.

\nh Segal, M., 1985. A potent transient outward current 
regulates excitability of dorsal raphe neurons.  
Brain Res 359,  347-350.

\nh Steinbusch, H.W., Nieuwenhuys, R., Verhofstad, A.A.,
 Van der Kooy, D., 1981. The nucleus raphe dorsalis 
of the rat and its projection upon the caudatoputamen. 
A combined cytoarchitectonic, immunohistochemical
 and retrograde transport study. J Physiol (Paris)
77, 157-174.

\nh Traub, R.D., Wong,  R.K.S., Miles, R., Michelson, H., 1991.  
A model of a CA3 hippocampal pyramidal neuron incorporating
voltage-clamp data on intrinsic conductances.  J Neurophysiol 66, 635-650. 

\nh Traub, R.D., Buhl, E.H., Gloveli, T., Whittington, M.A., 2003. 
Fast rhythmic bursting can be induced in layer 2/3 cortical
neurons by enhancing persistent Na$^+$ conductance
or by blocking BK channels.
 J Neurophysiol 89,  909-921.

\nh Williams, G.S.B., Smith,  G.D., Sobie, E.A., Jafri, M.S., 2010. 
Models of cardiac excitation-contraction coupling in ventricular myocytes.
Math Biosci 226,  1-15.

\nh  Xu, J., Clancy, C.E., 2008.  Ionic mechanisms of endogenous bursting in CA3
hippocampal pyramidal neurons: a model study. PLoS ONE 3, e2056.

\nh Yao, Y., Bergold, P., Penington, N.J., 2010.
Acute \CAN-dependent desensitization of 5-HT1A receptors
 is mediated by activation
 of protein kinase A (PKA) in rat serotonergic neurons. Neuroscience 169,
 87-97.

\nh Yamamoto, M., Steinbusch, H.W., Jessell, T.M., 1981. 
Differentiated properties of identified serotonin neurons in 
dissociated cultures of embryonic rat brain stem.
 J Cell Biol 91,142-152. 

\nh Yu, Y., Shu, Y., McCormick, D.A., 2008.
Cortical action potential backpropagation explains spike
threshold variability and rapid-onset kinetics. J Neurosci 28, 7260-7272.

 \end{document}